\documentclass{article}%
\usepackage{graphicx}
\usepackage{amsmath}
\usepackage{amsfonts}
\usepackage{amssymb}%
\newtheorem{theorem}{Theorem}

\begin{document}

\title{Area density of localization-entropy I: \\the case of wedge-localization}
\author{Bert Schroer\\CBPF, Rua Dr.\ Xavier Sigaud 150 \\22290-180 Rio de Janeiro, Brazil\\and Institut f\"ur Theoretische Physik der FU Berlin, Germany}
\date{updated June 2006}
\maketitle

\begin{abstract}
Using an appropriately formulated holographic lightfront projection, we derive
an area law for the localization-entropy caused by vacuum polarization on the
horizon of a wedge region. Its area density has a simple kinematic relation to
the volume extensive heat bath entropy of the lightfront algebra. Apart from a
change of parametrization, the infinite lightlike length contribution to the
lightfront volume factor corresponds to the short-distance divergence of the
area density of the localization entropy. This correspondence is a consequence
of the conformal invariance of the lightfront holography combined with the
well-known fact that conformality relates short to long distances. In the
explicit calculation of the strength factor we use the temperature duality
relation of rational chiral theories whose derivation will be briefly
reviewed. We comment on the potential relevance for the understanding of Black
hole entropy.

\end{abstract}

\section{Introduction, history description of main new result}

\textbf{Vacuum fluctuations and localization entropy in the historical
perspective} \smallskip

Localization entropy is a thermal manifestation of vacuum polarization;
different from the standard heat bath entropy of classical statistical systems
it is of purely quantum-physical origin. As the prize of quantum mechanical
(``Born''-) localization is paid by an uncertainty in momentum, localization
in causal relativistic QFT (``modular localization'' \cite{M-S-Y}) is always
accompanied by a thermal manifestation of the vacuum polarization at the
localization boundary. Whereas uncertainty relations involving the position
operator were identified as the characteristic properties of QM shortly after
its discovery, the understanding of the thermal signature of vacuum
fluctuations as a characteristic property of local quantum physics\footnote{In
the spirit of Haag's book \cite{Haag} we prefer the term \textit{local quantum
physics (LQP)} or \textit{algebraic QFT} (AQFT) whenever we want to
de-emphasize the use of field coordinatizations in favor of a more intrinsic
local operator-algebraic presentation of QFT.} is a recent observation. The
reason why this took such a long time is two-fold. On the one hand the thermal
properties are hard to see in the standard Lagrangian formulation, and on the
other hand the associated Hamiltonian is almost never directly experimentally
realizable, i.e.\ the conjugate time and the associated temperature can almost
never be associated with that of an actual observer. The only potential
exceptions are the Unruh Gedankenexperiment and the physics of black holes as
first observed by Hawking. In the latter case it is the curvature of spacetime
which creates the conditions for the timelike Killing symmetry outside a black
hole to be associated with the Hamiltonian translation of localization
thermality caused by vacuum polarization (or vacuum entanglement).

Vacuum fluctuations as an unavoidable attribute of local quantum physics were
first noted by Heisenberg \cite{Hei} when he computed (what we nowadays would
call) a \textquotedblleft partial charge\textquotedblright\ by integrating the
Wick-ordered zero component of a bilinear conserved current density over a
finite spatial volume. Heisenberg noticed that the current conservation law
does not control the infinitely strong particle-antiparticle vacuum
polarization at the boundary of the volume. Later conceptual and mathematical
refinements of QFT\footnote{These refinements resulted from a better
understanding of the nature of fields as operator-valued distribution,
requiring a smoothing in the definition of partial charges by a test function
which includes a compact smearing in time.} showed that these fluctuations can
be kept finite by avoiding sharp localization by allowing a ``halo'', i.e.\ a
region of ``fuzzy'' localization in a surface of finite thickness; in the
standard test function setting this corresponds to a test function decrease in
spacelike direction which interpolates between the unit strength and the value
zero \cite{Sw}. The dependence on the specific smoothing chosen to define the
``partial charge'' disappears in the infinite volume (thermodynamic) limit,
and the partial charge converges weakly against the unique global conserved charge.

In the presence of interactions, the quadratic vacuum fluctuations
(particle-antiparticle pairs) of interaction-free partial charges change into
vacuum polarization ``clouds'' involving an infinite number of
particle-antiparticle pairs. This phenomenon, which in the perturbative
context (where the number of pairs increases with the order of perturbation)
was first noticed with some surprise by Furry and Oppenheimer \cite{Fu-Op},
showed the limitation of Dirac's particle-based view (the hole
theory\footnote{The relativistic particle interpretation of quantum field
theory was finally abandoned when it became clear that Dirac's hole theory
although successful in low orders (see Heitler's book) cannot cope with
renormalization.}) and favored Jordan's more radical field-quantization
description of particle physics. But even though vacuum polarization as a
result of localization is an accepted fact, one finds often relapses into a
static kind of quantum mechanical picture where certain momentum space levels
are occupied by particles, and the entropy of a state is determined by
counting occupied levels; some authors even call QFT relativistic ``quantum
mechanics''. To be more explicit on this point, there exists a consistent
relativistic QM \cite{C-P}\cite{Bert}, but it lacks the vacuum polarization
and only possesses the standard probabilistic quantum mechanical Born
localization (in the relativistic particle context called Newton-Wigner
localization) which does not have a thermal manifestation. For a recent review
of Born-localization of particles and modular localization of fields
(including historical remarks about this physically rich topic) the reader is
referred to \cite{M-S-Y}.

Placed into a modern conceptual setting, the vacuum-polarizability through
modular localization can be backed up by a powerful theorem stating that the
existence of a so-called PFG localized in a region smaller than a
wedge\footnote{A general wedge $W$ is a Poincar\'{e} transform of the standard
wedge $W_{0}=\{x_{1}>\left\vert x_{0}\right\vert ,x_{2,3}\;
\hbox{arbitrary}\}$, and a subwedge region $\mathcal{O}$ is any region which
can be enclosed in a wedge $W\supset\mathcal{O}$. Wedges and their
intersections play a prominent role in formulations of QFT which trade
singular fields against algebras of (bounded) localized operators, since they
are the natural Poincar\'{e}-invariant families of localization regions
(invariant as families, not as individual algebras).} forces the theory to be
free. PFG is the acronym for vacuum \textbf{p}olarization-\textbf{f}ree
\textbf{g}enerator, which is an operator whose application to the vacuum
vector creates a one-particle state without any admixture of vacuum
polarization.
This version of the Jost-Schroer theorem \cite{S-W} (which was based on the
use of covariant fields) is a generalizion to the algebraic formulation of QFT (AQFT).

In any QFT in which the one-particle mass shell is separated from the
continuous part of the spectrum (mass gap assumption), there always exist PFG
operators affiliated with the global algebra $\mathcal{A}$ \cite{Haag}.
In the presence of interactions, the wedge regions are the smallest causally
complete region in which this continues to hold. The wedge region therefore
offers the best compromise between local algebras corresponding to the
interacting field concept and Wigner particle states \cite{S1}.
Wedge-localized operator algebras also turn out to present the simplest
``theoretical laboratory'' for understanding the thermal manifestations of localization.

These results have been established with the help of modular localization
\cite{BBS}. Modular localization is not a new physical principle of QFT, but
only stands for the inherent concept of localization in QFT based on the
commutativity of algebras. Localization is thus detached from the nominal
arguments of quantum fields $A(x)$ (i.e., ``field coordinatizations'' of
quantum observables) \cite{M-S-Y}. It draws its name from the mathematical
setting of the Tomita-Takesaki modular theory of operator algebras (see the
appendix), which happens to present the right concepts for achieving this separation.

The central issue of this paper is the observation that vacuum polarization on
individual localized operators can be related to collective properties of
causally localized operator algebras to which they are affiliated, and as a
consequence can be associated with an entropy. In particular it will be shown
that the area proportionality of localization-entropy is a generic property of
local quantum matter.

The idea that localized algebras may exhibit thermal properties was not known
at the time of the observations of Heisenberg and Furry-Oppenheimer; it came
from two different more recent sources. There is the famous physically
well-motivated observation by Hawking on thermal radiation of quantum matter
enclosed in a Schwarzschild black hole \cite{Haw}. Closely related is Unruh's
Gedankenexperiment \cite{Un} involving a uniformly accelerated observer whose
world-line is restricted to a Rindler wedge in Minkowski spacetime. In this
case the thermal manifestations of vacuum fluctuations are detached from the
presence of strong curvature effects of general relativity. Independently
there is the structural observation by Bisognano and Wichmann \cite{Bi-Wi}
which established that the restriction of the global vacuum state to a
wedge-localized subalgebra becomes a thermal KMS (Kubo-Martin-Schwinger) state
for arbitrary interacting matter content of the QFT. In fact they found this
thermal manifestation as a side result of their application of the modular
Tomita-Takesaki theory of operator algebras (which was discovered a decade
before, with important independent contributions coming from physicists doing
quantum statistical mechanics directly in the thermodynamic limit
\cite{Haag}). The special feature of the wedge regions is that the associated
modular objects have a direct physical interpretation in terms of geometric
symmetries (wedge-preserving Lorentz boosts and TCP transformation into the
opposite wedge). The connection between the Hawking-Unruh and the
Bisognano-Wichmann thermal manifestation of vacuum polarization was first
pointed out by Sewell \cite{Se}. Some remarks about modular theory and vacuum
polarization can be found in the appendix.

Once it can be argued that Heisenberg's observation about vacuum polarization
(but now placed within the context of localized operator algebras) leads to
thermal manifestations, the surface nature of this local quantum phenomenon
suggests that the thermal aspect is different from that of the standard heat
bath situation. Nevertheless, the quantitative computation of localization
entropy makes use of a unitary equivalence of the QFT under consideration in
the vacuum state with an auxiliary system in heat bath thermal state.

\medskip

\textbf{Outline of the computation of localization entropy} \smallskip

By the Bisognano-Wichmann theorem, the vacuum state restricted to the algebra
of a wedge is a KMS state of temperature $\beta=2\pi$ with respect to the
one-parameter group of Lorentz boosts preserving the wedge. The naive idea to
compute the associated entropy by the standard von Neumann formula from the
Gibbs partition function $\mathrm{Tr}\,\exp-\beta K$ (where $K$ is the
generator of the boosts) clearly fails, because $\exp-\beta K$ is not a trace
class operator, reflecting the infinite size of the system. In order to
control the divergence and relate it to the extension of the system and to the
sharp localization, we proceed in two steps.

The first step is the observation that the algebra of observables in a wedge
$W$ coincides with the algebra of observables localized in the (upper $=$
future) horizon $H(W)$ of the wedge
\begin{align}
\mathcal{A}(W)=\mathcal{A}(H(W)).
\end{align}
One may regard the horizon $H(W)\sim\mathbb{R}^{2}\times\mathbb{R}_{+}$ as a
subset of the associated lightfront $\mathbb{R}^{2}\times\mathbb{R}$. The
lightlike translations and the Lorentz boosts of the theory act on the
lightfront like the translations and dilations of the factor $\mathbb{R}$.
This symmetry extends by modular theory to the M\"obius group, thus defining
an associated M\"obius covariant net of algebras of observables localized in
subsets $A\times I$ of the lightfront ($A\subset\mathbb{R}^{2}$,
$I\subset\mathbb{R}$ intervals), such that
\begin{align}
\mathcal{A}_{\mathrm{LF}}(\mathbb{R}^{2}\times(a,\infty)) = \mathcal{A}%
_{\mathrm{LF}}(H(W_{a}))= \mathcal{A}(W_{a})
\end{align}
for all wedges $W_{a}=T_{a}(W)$ obtained by a translation of $W$ in the
lightlike direction.

The crucial observation besides the conformal symmetry in the lightlike
direction, is that this lightfront QFT exhibits no vacuum fluctuations in the
transverse directions. For free fields, this fact can be directly read off the
two-point function which on the lightfront exhibits a $\delta$-function in the
transverse direction (``ultra-locality''). In the interacting case, one again
has to appeal to modular theory which guarantees that the algebras of
lightfront observables localized in $A_{1}\times\mathbb{R}_{+}$ and in
$A_{2}\times\mathbb{R}_{+}$ with $A_{1}$ and $A_{2}$ disjoint subsets of the
transverse space $\mathbb{R}^{2}$ are algebraically in a tensor product
position. This means that the localization entropy $S_{\mathrm{loc}}%
(A\times\mathbb{R}_{+})$ associated with the algebra $\mathcal{A}%
_{\mathrm{LF}}(A\times\mathbb{R}_{+})$ is additive with respect to disjoint
subsets $A$ of $\mathbb{R}^{2}$, and hence proportional to the area $\vert
A\vert$.

Splitting off the area factor, the resulting area density of the localization
entropy is interpreted as the entropy of localization in $\mathbb{R}_{+}$ of a
chiral conformal QFT, defined by the M\"obius covariance of $\mathcal{A}%
_{\mathrm{LF}}$ in the lightlike direction.

The latter entropy still diverges. As we shall see, the divergence is due to
the sharp boundary of the localization region. In order to control it, the
second step is the observation (again by modular theory \cite{S-W}) that in
chiral CFT the algebra of the halfline $\mathbb{R}_{+}$ in the vacuum
representation is unitarily equivalent to the global algebra $\mathcal{A}%
_{\mathrm{CFT}}(\mathbb{R})$ in a thermal KMS state with inverse temperature
$\beta=2\pi$ with respect to the translations, and this equivalence extends to
subalgebras associated with intervals such that
\begin{align}
\pi_{0}(\mathcal{A}_{\mathrm{CFT}}(I)) = \pi_{\mathrm{KMS}}(\mathcal{A}%
_{\mathrm{CFT}}(\log I)).
\end{align}
Here, $\log I = (\ln x,\ln y)\subset\mathbb{R} $ if $I=(x,y)\subset
\mathbb{R}_{+}$.

For finite intervals $I\subset\mathbb{R}_{+}$, the entropy associated with
$\mathcal{A}_{\mathrm{CFT}}(\log I)$ in the KMS state can be computed by
standard methods of heat-bath thermality, by putting the system in a ``box''
of size $L=\vert\log I\vert$. As expected, it turns out to be finite and
proportional to $L$ for large $L$. The origin of the divergence of the
localization entropy is thus traced back to the sharp localization in
$\mathbb{R}_{+}$: by approximating $\mathbb{R}_{+}$ from the inside by an
interval $(x,y)$ with $x/y=\varepsilon$ in the limit $\varepsilon\to0$, the
entropy becomes proportional to $L=\vert\log I\vert= \vert\ln\varepsilon\vert
$. The parameter $\varepsilon$ here plays the role of the ``halo'' mentioned
before, related to the smoothing of the partial charge.

\medskip\textbf{Physical discussion of results} \smallskip

Combining the area law and the short-distance behavior at the boundary of the
localization region, the result of the previous discussion can be cast in the
formula for the localization entropy
\begin{align}
\label{area}S_{\mathrm{loc}}(A\times\mathbb{R}_{+}) =\vert A\vert\cdot
\lim_{\varepsilon\to0} \frac{c}{12}\vert\ln\varepsilon\vert
\end{align}
if $\mathbb{R}_{+}$ is approximated from the inside by an interval $(x,y)$
such that $x/y=\varepsilon$.

This is the seeked for area law for entropy of wedge localization expressed in
terms of the geometric holographic data. The holographic matter-dependent
parameter $c$ measures the degrees of freedom of the associated extended
chiral theory, and is in typical cases related to the Virasoro algebra
constant\footnote{Unlike the chiral components from the chiral decomposition
of $d=1+1$ CFT, the chiral structure emerging from holographic projection
comes a priori without an energy momentum tensor (even if the bulk theory had
one).} (the reason for the chosen notation). The long distance part $L$ in the
standard heat bath volume divergence on the lightfront has been transformed
into a short-distance $\varepsilon\rightarrow0$ divergence. Actually, by
conformal covariance, if any other interval than $\mathbb{R}_{+}$ is
approximated from the inside by a smaller interval with a short-distance
\textquotedblleft halo\textquotedblright, then $\varepsilon$ is the conformal
cross-ratio of the four points defining the pair of intervals.

Note that the localization entropy is described by a one-parameter family of
approximating Gibbs states labeled by $\varepsilon$; all members of this
family have the same $\beta=2\pi$ ``Hawking temperature''\footnote{The fact
that a value of $\varepsilon$ of the order of the Planck length brings
localization entropy down into the range of the Bekenstein entropy does
unfortunately not reveal anything about their relation (as long as one does
not know something about their coupling).}. As in the case of the
thermodynamic limit only the leading behavior for $\varepsilon\rightarrow0$ is
expected to be universal, while the nonleading corrections depend on the
physical nature of the boundary.
In particular, the leading singular behavior cannot be used to probe some
suspected unknown short distance behavior of the bulk theory. In fact it does
not even depend on whether the bulk algebra permits a description in terms of
singular pointlike generating covariant fields as long as the spacetime
symmetries are valid and certain intersections of wedge algebras are
nontrivial. In a continuation of this work it will be shown that the area
behavior of localization thermality continues to hold for other geometries of
causally closed localization regions as double cones. Although the present
analysis is limited to localization in Minkowski spacetime, there is no reason
to doubt that the area behavior of localized quantum matter carries over to
QFT in curved spacetime. But some of the concepts need adjustments, especially
if one is dealing with interacting quantum matter. Instead of the Poincar\'{e}
symmetry one expects that the newly discovered local covariance principle (and
the resulting more subtle operator implementation of local diffeomorphism
placed within the setting of modular theory) will play a crucial role.

Different from the Bekenstein area law which holds in classical field theories
with special geometric properties (Einstein-Hilbert-like field theories), the
area law for the entropy caused by quantum localization is a \textit{general
property of matter in local quantum physics}. Whereas the localization entropy
is a generic manifestation of local quantum matter, the Bekenstein law has
been interpreted as the quantum entropy carried by the gravitational degrees
of freedom in a future quantum gravity. Since also the Bekenstein law is
claimed to be related to microscopic statistical mechanics aspects of Hawking
radiation, the (conceptually different) issue of localization entropy
superficially complicates\footnote{What is being measured according to
Hawking's calculation is radiation of quantum matter and not radiation of
gravitons.} the use of Bekenstein's observation as a means to get a hold on
quantum gravity, although in the long run it may turn out to be essential for
the solution of the quantum gravity problem. More remarks can be found in the
concluding remarks.

The presentation of the correspondence between holographically projected heat
bath- and localization-thermality reveals that the abstract reason for the
divergences is the same, namely in both cases one is approximating a copy of
the \textit{``monade''} (the unique hyperfinite type III$_{1}$ von Neumann
factor algebra \cite{Su} as classified in the work of Connes, see appendix).
\textit{This algebra already comes with properties which preempt the (modular)
sharp localization and thermal behavior} (in fact, it possesses no autonomous
pure states at all)\footnote{No other algebra than the monade can account for
the localization and thermal properties of QFTs, although other ``large
enough'' algebras as the type I$_{\infty}$ algebras contain monades.}. As a
result of the identity of the operator algebra of the wedge $\mathcal{A}(W)$
with its (upper) horizon algebra $\mathcal{A}_{\mathrm{LF}}(H(W))$,
the entropy divergence law for the conformal holographic projection is the
same as for the wedge algebra. As a result of the absence of a mass scale the
entropy is strictly speaking only determined up to a (common for different
quantum matter) scale factor.

Sequences of ordinary quantum mechanical type I$_{\infty}$ operator algebras
which are inclusive and converge towards a limiting algebra are called
\textit{funnels}. In our case we are interested in funnels which approximate
the monade algebra. The ideal situation for the conceptual setting of the
thermodynamic limit discussion would be such a funnel sequence which is also
inclusive in the spacetime sense, i.e.\ which exhausts the Minkowski spacetime
in the limit. Since both the algebra of an open system (infinite volume) in a
KMS state of finite temperature with respect to the standard Hamiltonian, as
well as an algebra of sharply localized observables in the vacuum state
representation are monade algebras, there should be an analogy in their funnel
approximations. Our result above shows that if we work with the holographic
projection instead of the bulk matter, the analogy becomes an equivalence.

Although, as will be seen, the mathematical setting of modular localization
permits clear definitions and rigorous derivations, one faces serious problems
when one tries to convert the thermal manifestations of localization into
observational consequences. These difficulties also explain why it took such a
long time after having noticed the presence of localization-caused vacuum
polarization to become also aware of their thermal consequences. In order to
explain this important point in some more detail, let us first recall that the
notion of temperature in the standard heat bath setting of statistical
mechanics is related to the time translation and the corresponding standard
Hamiltonian. In terms of this Hamiltonian one defines a finite volume tracial
Gibbs state at inverse temperature $\beta$. To arrive at thermodynamic
equilibrium in which the boundary effects become insignificant, one performs
the thermodynamic infinite volume limit in which the appropriately normalized
Gibbs state converges towards a KMS state associated with the Hamiltonian
automorphism. Independent of any details, KMS states associated with a
Hamiltonian $H$ are known to fulfill an abstract form of the second law of
thermodynamcs \cite{Pu-Wo} which can be expressed in terms of the following
inequality
\begin{align}
\label{2law}E_{H} \equiv\left\langle U\Omega_{\beta}\left\vert H\right\vert
U\Omega_{\beta}\right\rangle  &  \geq\left\langle U\Omega_{\beta}\left\vert
1-e^{-H}\right\vert U\Omega_{\beta}\right\rangle =\nonumber\\
&  =\left\langle U\Omega_{\beta}\vert U\Omega_{\beta}\right\rangle
-\left\langle U^{\ast}\Omega_{\beta}\vert U^{\ast} \Omega_{\beta}\right\rangle
= 0
\end{align}
which is a consequence of the KMS property written as
\begin{align}
\left\langle A\Omega_{\beta}\vert e^{-H}\vert B\Omega_{\beta}\right\rangle
=\left\langle B^{\ast}\Omega_{\beta}\vert A^{\ast}\Omega_{\beta}\right\rangle
,\,\,A,B\in\mathcal{A}.\nonumber
\end{align}
Here $U$ denotes any unitary operator in the global observable algebra whose
application to the thermal state represents the change caused by an external
force which acts during a finite time. The inequality (\ref{2law}) expresses
the impossibility to extract energy ($E_{H}<0$) without causing a permanent
change of the external conditions (impossibility of a perpetuum mobile).
Standard assumptions about the form of the Hamiltonian allow to convert this
abstract form of the second thermodynamic law into the more concrete
quantified form in terms of an entropy function.

Modular theory permits to repeat these arguments word for word in case the
operator algebra is a localized algebra $\mathcal{A(O)}$ and $\Omega_{\beta}$
is any vector on which this algebra acts in a cyclic and separating manner,
e.g., the vacuum state $\Omega=\Omega_{\infty}$ (if $\beta$ is the inverse
temperature). The modular substitute for $H$ is the generator $K$ of the
modular group $\Delta^{it}=e^{iKt}$ often referred to as the ``modular
Hamiltonian''. Although modular theory guarantees the existence of the modular
objects, it does not provide a physical interpretation of the modular
Hamiltonian $K$ and the modular ``time'' $t$. Only in the fortunate case in
which $\Delta^{it}$ admits a geometric interpretation one can think of a
Gedankenexperiment for observing the thermal consequences in terms of thermal
radiation. The only case in Minkowski space QFT is the wedge-localized
operator algebra relative to the vacuum state; in this case Bisognano and
Wichmann showed that the modular automorphism is the wedge-preserving
Lorentz-boost. In curved space time there are more possibilities to find
Killing symmetries which leave subregions invariant. In those cases it does
not matter whether this occurs in the context of general relativity or in
analog situations in acoustics, hydrodynamics or optics \cite{U-S} where such
situations are generated by encoding microscopic properties into an
\textquotedblleft effective\textquotedblright\ spacetime metric and the
resulting effective description is a quantum theory with a finite propagation
speed. Even if one does not share the optimism about its experimental
accessibility \cite{Visser}, the subject is of sufficient intrinsic
theoretical interest. Structural properties in QFT, which is the most
successful theory of all times up to now, are of paramount interest even if
they do not lead to experimentally testable results, as long as they lead to
new probes of the underlying principles.

The derivation of the area density formula (\ref{area}) which will be the main
topic of this paper is based on:

\begin{itemize}
\item The holographic\footnote{'t Hooft first used this terminology for a
speculative property which he expects to be a property of black holes in a
future quantum gravity theory. Here this terminology is used for a generic
property in QFT in curved spacetime, in particular standard QFT in Minkowski
spacetime.} projection of the wedge algebra onto a transversely extended
chiral algebra (the second section).

\item The possibility (limited to 2-dim. conformal theories) to pass from
localiza\-tion-caused thermal behavior to global heat bath thermality and vice
versa (section 3) and the existence of an ``intrinsic'' thermodynamic limit
sequence (a ``box'' which preserves the spacetime covariances) in conformal QFT.

\item In order to obtain concrete limiting formulas from the last fact one
uses again the chiral nature of the conformal theories which permits to use
asymptotic estimates of the Cardy-Verlinde type (section 4).
\end{itemize}

Most of these points have previously appeared in different contexts in
previous work of the author. They will be presented in a form adapted to the
present purpose.

\section{Reviewing lightfront holography and the consequences of absence of
transverse vacuum polarization}

The study of thermal aspects of localization is greatly simplified by using
instead of the original operator algebra its holographic projection. In case
of the wedge algebra, the holographic projection leads to the lightfront
algebra. The latter is related to what used to be called ``lightcone
quantization'', in fact it could be seen as a conceptual and mathematical
rescue operation to save some of the intuitive content of the latter while
avoiding the conceptual pitfalls of the too naive view of what constitutes
QFT. Different from the old approach it should not be viewed as a new
quantization, but rather as a different spacetime encoding of a given QFT. In
other words it is a concept which reprocesses the spacetime affiliation of the
algebraic substrate indexed in terms of spacetime regions in the ambient
space, to a radically different one in which subalgebras of the same global
algebra are indexed by localized regions on a lightfront. (This is a manifold
which contrary to the ambient manifold is neither globally nor even locally
hyperbolic) \cite{S1}.

If there are no interactions this can be done directly in terms of free
fields. One may first express the restriction of a free field $A(x)$ to the
wedge by parametrizing the latter as
\begin{align}
x= (x^{0}  &  =r\sinh\chi,x^{1}=r\cosh\chi,x_{\perp})\\
p= (p^{0}  &  =m_{\mathrm{eff}}\cosh\theta, p^{1}=m_{\mathrm{eff}}\sinh
\theta,p_{\perp})\nonumber
\end{align}
where the ``effective mass'' $m_{\mathrm{eff}}=\sqrt{p_{\perp}^{2}+m^{2}}$
depends on the transverse momentum. Then
\begin{align}
\label{free}A_{W}(r,\chi,x_{\perp})\equiv A(x)\vert_{W}=(2\pi)^{-\frac32}%
\int(e^{i(m_{\mathrm{eff}}r\cosh(\chi-\theta)- ip_{\perp}x_{\perp}}A^{\ast
}(p)+h.c.)d^{2}p_{\perp}\frac{d\theta}{2}%
\end{align}
where the creation and annihilation operators satisfy
\begin{align}
\left[  A(p),A^{\ast}(p^{\prime})\right]  =2p_{-}\delta(p_{-}-p_{-}^{\prime
})\delta(p_{\perp}-p_{\perp}^{\prime}),\qquad p_{-}=\frac12m_{\mathrm{eff}%
}(p_{\perp})e^{-\theta}%
\end{align}
implying the factorization of the one-particle Hilbert space $H_{1}%
=H_{-}\otimes H_{\perp}$ into lightlike and transverse states. The restriction
to the (upper) horizon $H(W)$ is done in terms of the limit $r\rightarrow0$,
$\chi\sim\left\vert \ln r\right\vert \rightarrow\infty$ such that
$x_{+}=re^{\chi}$ remains finite and $x_{-}=re^{-\chi}\rightarrow0.$ The
resulting expression for the limiting singular operator is
\begin{align}
\label{lf}A_{H(W)}(x_{+},x_{\perp})=(2\pi)^{-\frac{3}{2}}\int_{0}^{\infty
}\frac{dp_{-}}{2p_{-}} \int\left(  A^{\ast}(p)e^{ip_{-}x_{+}}e^{-ip_{\perp
}x_{\perp}}+h.c.\right)  d^{2}p_{\perp}.
\end{align}
In order to see that the change of variable $p_{-}=\frac12 m_{\mathrm{eff}%
}e^{-\theta}$ with the $p_{\perp}$ dependent pre-factor is allowed, one has to
remember that this change does not modify the operator after integrating with
the relevant limiting class of smearing functions which vanish at the origin
$p_{-}=0$\footnote{This is well-known for the zero mass scalar free field in
two dimensions. In that case the exponential contains an engineering
dimension-setting mass parameter which has no bearing on the energy-momentum
spectrum and which drops out after test function smearing within the
appropriate space of functions \cite{S1}.}. The terminology ``lightfront
restriction'' therefore only agrees with its naive geometric meaning $x_{-}=0$
in this mass shell representation, while doing this in correlation functions
would give nonsensical results.

The linear extension of the horizon yields the lightfront. The formula for the
lightfront fields $A_{\mathrm{LF}}(x_{+},x_{\perp})$ extends that for
$A_{H(W)}$ from $x_{+}>0$ to all $x_{+}$.

From (\ref{lf}) we obtain the two-point and commutator functions
\begin{align}
\left\langle A_{H(W)}(x_{+},x_{\perp})A_{H(W)}(x_{+}^{\prime},x_{\perp
}^{\prime})\right\rangle _{0}  &  =\delta(x_{\perp}-x_{\perp}^{\prime}%
)\cdot\int_{0}^{\infty}\frac{dp_{-}}{4\pi p_{-}} e^{-ip_{-}(x_{+}%
-x_{+}^{\prime})}\nonumber\\
\left[  A_{H(W)}(x_{+},x_{\perp}),A_{H(W)}(x_{+}^{\prime},x_{\perp}^{\prime
})\right]   &  =\delta(x_{\perp}-x_{\perp}^{\prime})\cdot\frac1{4i}%
\varepsilon(x_{+}-x_{+}^{\prime})\nonumber\\
\left[  \partial_{+}A_{H(W)}(x_{+},x_{\perp}),\partial_{+}A_{H(W)}%
(x_{+}^{\prime},x_{\perp}^{\prime})\right]   &  = \delta(x_{\perp}-x_{\perp
}^{\prime})\cdot\frac i2\delta^{\prime}(x_{+} -x_{+}^{\prime}) \label{lpr}%
\end{align}
The crucial point here is the $\delta$-function in the transverse variables,
meaning the absence of transverse vacuum fluctuations. The nontrivial task is
to show that this feature, which is the origin of the area law for the
localization entropy (see below), survives in the presence of interactions.

If one wants unrestricted test function spaces on the lightfront one should
start from the derivative field $\partial_{+}A$ which creates the same Hilbert
space as $A.$ By a process called Haag-dualization the local operator algebras
generated by $\partial_{+}A$ are known to be the same as those generated by
$A$ (this is similar to \cite{GLW}). The lightfront restriction of the
derivative field has the algebraic structure of a transversely extended chiral
theory for an abelian current (\ref{lpr}) where the $\delta^{\prime}$-function
represents the chiral aspect.

In the presence of interactions the lightfront restriction suffers from the
same problem as the derivation of equal time canonical commutation relations;
the obstacle in both cases is the infinite wave function renormalization (the
divergence of the integral over the Kallen-Lehmann spectral function)
\cite{S1}.

The analog of the mass-shell representation for interacting fields is fairly
involved since it requires the apparatus of LSZ scattering theory. The latter
leads to the following so-called Glaser-Lehmann-Zimmermann expansion of
interacting Heisenberg fields in terms of incoming free fields
\[
A(x)=\sum\frac{1}{n!}\int_{H_{m}}\cdots\int_{H_{m}}e^{i\sum_{i=1}^{n}%
p_{i}x_{i}}a(p_{1},\ldots,p_{n}):A_{in}(p_{1})\cdots A_{in}(p_{n}):\prod
_{i}\frac{d^{3}p_{i}}{2p_{i0}}%
\]
which on the lightfront reduces to
\begin{align}
&  A_{\mathrm{LF}}(x_{+},x_{\perp})=\sum\frac{1}{n!}\int_{H_{m}}\cdots
\int_{H_{m}}e^{i\sum_{i=1}^{n}(p_{i-}x_{i+}-p_{i\perp}x_{i\perp})}%
a(p_{1},\ldots,p_{n})\times\nonumber\\
&  \hskip40mm\times:A_{in}(p_{1})\ldots A_{in}(p_{n}):\prod_{i}\frac{dp_{i-}%
}{p_{i-}}d^{2}p_{i\perp}%
\end{align}
Here the coefficient functions are mass shell restrictions of Fourier
transforms of retarded correlation functions, and the integration region
$H_{m}$ consists of the forward and backward mass shell $H_{m}=H_{m}^{+}\cup
H_{m}^{-}$. Besides being extremely formal (the convergence properties of such
representations are unknown, it is nothing more than a collection of
formfactors of $A$), the use of formulas based on scattering theory would
defeat the whole motivation for the \textit{lightfront formalism} which was to
\textit{simplify certain aspects of the original dynamical problem}. A
different spacetime encoding cannot accomplish dynamical miracles; the best
one can hope for is that certain aspects of vacuum polarization and their
thermal manifestation which one is interested in, become simpler (naturally at
the prize of other properties, in this case of more complicated description of
particle and S-matrix aspects).

A conceptually and mathematically superior approach consists in avoiding field
coordinatizations altogether in favor of the modular localization formalism of
operator algebras. In the presence of interactions this is in fact the only
approach to holography.

The prototype situation of an operator algebraic approach in this paper will
be that of a Rindler-Unruh \cite{Un} wedge algebra whose holographic
projection is the (upper) causal horizon which covers half a lightfront. The
starting point is the equality of the wedge algebra with its holographic
projection
\begin{equation}
\mathcal{A}(W)=\mathcal{A}(H(W))\equiv\mathcal{A}_{\mathrm{LF}}(H(W))
\end{equation}
which in the absence of interactions follows from our free field computations;
in general it is considered as part of the definition of what constitutes a
causal and local quantum field theory\footnote{It is the limiting case of the
``causal shadow property'' of spacelike surfaces.}, i.e.\ it belongs to those
structural properties which remain unaffected by interactions. Although the
wedge algebra is equal to that of its lightfront horizon, this does not apply
to the localized substructures; in fact it is just the simpler spacetime
localization and vacuum polarization aspects of the right hand side which
facilitates greatly the computation of certain quantities as the entropy.

The localization substructure along the lightray is obtained by taking
intersections of algebras $\mathcal{A}_{\mathrm{LF}}(H(W_{a}))$. Studying
their modular groups, in several investigations it has been noted that the
resulting structure is that of a conformal chiral AQFT in the longitudinal
direction. In particular the wedge-preserving Lorentz boosts which are the
modular groups of the wedge algebras $\mathcal{A}_{\mathrm{LF}}(H(W_{a}))$
relative to the vacuum acts as dilations of the intervals $(a,\infty
)\subset\mathbb{R}_{+}$ on the lightfront \cite{Se}\cite{Su-Ve}%
\cite{Gu-Lo-Ro-Ve}, and their differences yield the generator of the
translations. Together with the modular groups of relative commutants
$\mathcal{A}_{\mathrm{LF}}(H(W_{a}))^{\prime}\cap\mathcal{A}_{\mathrm{LF}%
}(H(W_{a}))$, one obtains the M\"obius group. In contrast, the local
resolution in the transverse directions (i.e.\ the directions along the edge
of the wedge) is the result of more recent investigations \cite{S2}. These
results show in particular that the holographic lightfront projection has no
transverse vacuum polarization, a fact which is related to the \textit{radical
change of the spacetime interpretation in the re-processing of the ambient
algebraic substrate to its holographic projection.} In other words the
holographic projection leads to a system which behaves as a transverse quantum
mechanics at a fixed time; the vacuum state exhibits fluctuations only in the
lightlike direction.

In algebraic terms, the absence of transverse vacuum fluctuations means that
the global lightfront algebra tensor factorizes under ``transverse
subdivisions'':
\begin{align}
\mathcal{A}_{\mathrm{LF}}(\mathbb{R}^{2}\times\mathbb{R}_{+})\cong
\mathcal{A}_{\mathrm{LF}}(A\times\mathbb{R}_{+})\otimes\mathcal{A}%
_{\mathrm{LF}}(A^{\prime}\times\mathbb{R}_{+})
\end{align}
where $A\subset\mathbb{R}^{2}$ and $A^{\prime}=\mathbb{R}^{2}\setminus A$, and
this factorization is inherited by subalgebras associated with intervals
$I\subset\mathbb{R}_{+}$ in the lightlike direction (see below). Although a
detailed derivation of the localization structure on the horizon of the wedge
requires a substantial use of theorems about modular inclusions and
intersections (for which we refer to \cite{S2}\cite{S4}), the tensor
factorization of the horizon algebra relies only on the following structural
theorem in operator algebras\footnote{This is not the only argument for the
absence of transverse fluctuations (for other proofs see \cite{S2}) but it is
the most general one.}:

\begin{theorem}
(Takesaki \cite{Tak}) Let $\left(  \mathcal{B},\Omega\right)  $ be a von
Neumann algebra with a cyclic and separating vector $\Omega$ and
$\Delta_{\mathcal{B}}^{it}$ its modular group. Let $\mathcal{A}\subset
\mathcal{B}$ be an inclusion of two von Neumann algebras such that the modular
group $Ad\Delta_{\mathcal{B}}^{it}$ leaves $\mathcal{A}$ invariant. Then the
modular objects of $\left(  \mathcal{B},\Omega\right)  $ restrict to those of
$\left(  \mathcal{A}e_{\mathcal{A}},\Omega\right)  $ where $e_{A}$ is the
projection $e_{\mathcal{A}}H=\overline{\mathcal{A}\Omega}$ as well as to those
of $\left(  \mathcal{C}e_{\mathcal{C}},\Omega\right)  $ with $\mathcal{C}%
=\mathcal{A}^{\prime}\cap\mathcal{B}$ the relative commutant of $\mathcal{A}$
in $\mathcal{B}$ and $e_{\mathcal{C}}H=\overline{\mathcal{C}\Omega}.$
Furthermore the algebra $\mathcal{A} \vee\mathcal{C}$ is unitarily equivalent
to the tensor product $\mathcal{A}\otimes\mathcal{C}$ in the tensor product
Hilbert space.
\end{theorem}

In the application to lightfront holography we choose $\mathcal{B}%
=\mathcal{A}(W) \equiv\mathcal{A}_{\mathrm{LF}}(\mathbb{R}^{2}\times
\mathbb{R}_{+})$. Its modular group is the Lorentz boost $\Lambda_{W}(-2\pi
t)$ which in the holographic projection becomes a dilation. The dilation
invariance of the algebra $\mathcal{A}= \mathcal{A}_{\mathrm{LF}}%
(A\times\mathbb{R}_{+})$ is geometrically obvious. The relative commutant is
$\mathcal{C}=\mathcal{A}_{\mathrm{LF}}(A^{\prime}\times\mathbb{R}_{+})$. Thus
because entropy is additive under tensor products, and the subdivisions can be
continued to finer partitions, it follows for the localization entropy
\begin{align}
S_{\mathrm{loc}}(A_{1}\cup A_{2}\times\mathbb{R}_{+})= S_{\mathrm{loc}}%
(A_{1}\times\mathbb{R}_{+}) + S_{\mathrm{loc}}(A_{2}\times\mathbb{R}_{+})
\end{align}
whenever $A_{1}$ and $A_{2}$ are disjoint subsets of $\mathbb{R}^{2}$. Since
the area is the only additive invariant of subsets of $\mathbb{R}^{2}$, we get
the area law
\begin{align}
S_{\mathrm{loc}}(A\times\mathbb{R}_{+})\propto\vert A\vert.
\end{align}

The area density of entropy is the localization entropy of an auxiliary chiral
theory on the lightray given by the algebras $\mathcal{A}_{\mathrm{CFT}%
}(I)=\mathcal{A}_{\mathrm{LF}}(A\times I)$ with a unit area $A$. Whereas the
area behavior of localization entropy was known \cite{S2}, the actual
computation for a chiral localization interval remained an open problem which
we will address in the sequel.

There is another way (also based on modular operator algebra theory) by which
the transverse tensor factorization can be obtained \cite{S4}; it uses the
lightlike energy positivity and cluster factorization in the vacuum state. The
interesting aspect of the above theorem is that it does not use explicitly the
vacuum properties so that it could be useful in generalizing the thermal
manifestations of fluctuations near the causal horizon to global KMS states
(not considered in this paper).

Note that the conformal invariant chiral structure along the lightray does not
imply that the ambient theory is massless. Whereas the short-distance limit
(leading to critical universality classes) changes the theory, the holographic
projection takes place in the same Hilbert space as the ambient theory since
the particle creation/annihilation operators of the massive particles and the
representation of the Poincar\'{e} group have not changed; but the
interpretation in terms of localization of $\mathcal{A}_{\mathrm{LF}}$ is
radically different from that of $\mathcal{A},$ in particular these QFTs are
relatively nonlocal (which is linked to the fact that certain Poincar\'{e}
transformations, including the opposite lightray translation, act
non-geometrically on $\mathcal{A}_{\mathrm{LF}}$).

Different from the equal time canonical structure which breaks down for
interacting properly renormalizable fields, there is no such short distance
restriction on the generating field of the holographic projection; whereas the
short distance behavior of canonical fields must remain close to that of free
fields, fields on the lightray exist for arbitrary high anomalous dimensions.
The only problem is that one cannot get to those anomalous dimensional
lightfront fields by the above pedestrian restriction procedure based on the
mass shell representation. In view of the fact that lighfront holography
involves a very radical spacetime re-processing of the algebraic substrate,
this is not surprising. The dependence of the commutator on the transverse
coordinates $x_{\perp}$, encoded in the quantum mechanical derivative-free
$\delta$-function, is directly related to the transverse factorization of the
vacuum, i.e.\ to the factorization of the algebra of a cylinder (finite
transverse extension) in lightray direction into tensor products upon
subdivision into sub-cylinders. Any extensive quantity as an entropy, which
behaves additively for independent subsystems, is then additive in transverse
direction and hence follows an area law \cite{S2}\cite{S4}.

If one wants to translate the algebraic tensor factorization back into the
language of fields, then for consistency reasons the commutation relation of
the formal pointlike generators must be of the form\footnote{The field
generators of local transverse factorizing operator algebras must have the
claimed form of the spacetime commutation relations for reasons of
consistency; in particular the appearance of derivatives in the transverse
$\delta$-functions would destroy the factorization.}%
\begin{equation}
\left[  A_{\mathrm{LF}}(x_{+},x_{\perp}),B_{\mathrm{LF}}(x_{+},x_{\perp
}^{\prime})\right]  =\delta(x_{\perp}-x_{\perp}^{\prime})\sum_{n}\delta
^{(n)}(x_{+}-x_{+})C_{\mathrm{LF}n}(x_{+},x_{\perp})
\end{equation}
where the sum goes over a finite number of derivatives of $\delta$-functions
and $C_{\mathrm{LF}n}$ are (composite) operators of the model. The presence of
the quantum mechanical $\delta$-function and the absence of transverse
derivatives expresses the transverse tensor factorization of the vacuum,
i.e.\ all the field theoretic vacuum polarization has been compressed into the
$x_{+}$ lightray direction. As in QM, one can of course get derivatives of the
transverse $\delta$-function by using transverse derivatives of the generating
lightfront fields; the main point which secures the transverse tensor
factorization is that \textit{there exist generating fields without
derivatives}. In view of the fact that in chiral theories the existence of
pointlike field generators follows from the covariance structure of the
algebraic setting \cite{Joe}, there seems to be no problem to construct
pointlike fields in this transverse extended chiral theory along the same
lines; the local structure of the commutation relations is then a consequence
of locality and M\"{o}bius covariance \cite{S2}.

The absence of transverse vacuum fluctuations leads to the additivity of
entropy under transverse subdivisions, i.e.\ to the notion of area density of
entropy. The holography reduces the calculation of the area density to a
calculation of localization entropy of an associated chiral CFT on the
lightray. The basic problem, which will be addressed in the next section, is
how to assign a localization entropy to an interval on the (compactified) lightray.

If the absence of transverse vacuum fluctuations and the ensuing area behavior
would be limited to wedge-localization, the present conceptual setting would
not be so interesting. As a confidence-building extension one would like to
establish these facts at least for the compact causally closed double cone
region $\mathcal{D}$. In this case there is no geometric candidate for the
modular group of ($\mathcal{A(D)},\Omega$) when the underlying QFT is not
conformal invariant. For conformal covariant QFTs\footnote{$\mathcal{D}$ is in
fact conformally equivalent to $W$ \cite{Haag}.} on the other hand the modular
group consists of a one-parameter conformal subgroup which involves a chiral
M\"obius transformation with two fixed points in the radial variables $r_{\pm
}$ \cite{Hi-Lo}. There are convincing but not rigorous arguments \cite{S-Wi2}
to the effect that also in the massive case, the modular group close to the
boundary $\partial\mathcal{D}$ becomes asymptotically equal to the action of
this conformal group. The upper boundary $\partial D_{+}$ is a causal horizon
for $\mathcal{D}$ and the angular rotations would correspond to the transverse
translations on the wedge, i.e.\ to the directions which are free of vacuum
polarization. Though analogies are helpful, the double cone situation is
sufficiently different and warrants a separate presentation to which we hope
to return in a separate paper.

This algebraic structure of the commutation relations (\ref{lpr}) reveals
another interesting (and for our purpose important) information: the
lightfront projection places a new infinite dimensional symmetry group into
evidence which is of the Bondy-Metzner-Sachs type\footnote{The BMS group would
be associated to the automorphism strucure of commutation relations in which
the transverse $\delta$-function is replaced by the directional $\delta
$-function depending on two angles; a situation which one expects in case of
double cone holography \cite{S2}. This issue will be treated in a separate
part II to this paper.}. Such infinite-dimensional groups arose first in the
investigation of asymptotic behavior of zero mass theories and in particular
for asymptotically flat classical spacetimes (in the sense of Penrose). In our
local quantum physics setting they simply arise from transverse extensions of
the Diff($S^{1}$) symmetry of chiral theories%
\begin{align}
(x_{\perp,}x_{+})  &  \rightarrow(x_{\perp,}^{\prime}x_{+}^{\prime
})=(Ex_{\perp},\psi(x_{\perp,}x_{+}))\\
\psi(x_{\perp,}\cdot)  &  \in\mathrm{Diff}(S^{1})\nonumber
\end{align}
where $Ex_{\perp}$ is a Euclidean transformation in the transverse direction
and $x_{+}^{\prime}=\psi(x_{\perp,}x_{+})$ an $x_{\perp}$-dependent
diffeomorphisms of the compactified lightray coordinate $x_{+}\in
\overline{\mathbb{R}}\equiv S^{1}$. This symmetry of the algebraic structure
is unitarily implemented on the operator-algebraic level. It was already
present in the ambient setting but went unnoticed because it does not have the
form of a quantum Noether symmetry. This is because in the ambient bulk
setting it belongs to an infinite group of ``fuzzy'' symmetry transformations,
i.e.\ algebraic covariances similar to the localization preserving modular automorphism.

Since the issue of emergence of infinite symmetry groups in holographic
projections is of no direct importance for the thermal manifestations of
modular localization in this paper, the appearance of quantum B-M-S like
groups and their use in the quantum aspects of the conformal infinity in the
sense of Penrose will be deferred to a separate publication. Although the
Poincar\'{e} group continues to act on the lightfront operators, the
``visible''\ part consists only of a 7-parametric subgroup: the 3-parametric
subgroup in the wedge plane (the boost and two lightlike translations), the
3-parametric transverse Euclidean group, and the 1-parametric subgroup of the
Wigner little group given by the lightfront preserving lightlike translations
of the edge of the wedge.

\section{The interpretation of localization-caused thermality on the horizon
in terms of heat bath thermal behavior on the lightfront}

It has been known for some time that under special conditions the distinction
between heat bath and localization thermality becomes blurred. One such
situation has been studied in conformal two-dimensional models and named
appropriately ``Looking beyond the Thermal Horizon'' \cite{S-W} whereas a
similar situation in higher dimensions was presented as a ``converse
Hawking-Unruh effect''\cite{Gu-Lo}. The basic question is under what
circumstances a heat bath KMS state on a global operator algebra which is
associated to the translation automorphism may be interpreted as the
restriction of a state which is the vacuum on an appropriately constructed
extended global algebra (``behind the horizon''); in other words the thermal
KMS state on a given algebra is viewed as the restriction of a state which
happens to be the vacuum on a suitably extended algebra. In general the
commutant $\mathcal{A}^{\prime}$ of a global algebra $\mathcal{A}$ in a KMS
state is anti-isomorphic to the global algebra but there is no geometric
interpretation\footnote{The anti-isomorphism is used in the ``thermofield
formalism'' for a tensor product ``doubling'' which permits to use the Feynman
formalism in the thermal setting. But note that although the tensor
factorization holds for Gibbs states, it breaks down in the thermodynamic
limit algebra which is the monade algebra.}, it remains a ''shadow-world''. So
using a somewhat colorful terminology the question is: under what
circumstances can the abstract commutant $\mathcal{A}^{\prime}$ be viewed as
``virgin territory'' geometrically localized behind an causal horizon,
i.e.\ as the local algebra in the vacuum representation of a QFT on an
extended space $M_{\mathrm{ext}}$ into which $M$ is embedded, such that
\begin{align}
\mathcal{A}_{\mathrm{KMS}}(M)  &  =\mathcal{A}_{\mathrm{vac}}(M)\\
\mathcal{A}_{\mathrm{KMS}}(M)^{\prime}  &  =\mathcal{A}_{\mathrm{vac}%
}(M^{\prime})\nonumber
\end{align}
where $M^{\prime}$ is the causal disjoint of $M$ in $M_{\mathrm{ext}}$.

For our purpose it is sufficient to understand this for a translational KMS
state at temperature $\beta=2\pi$ of a chiral algebra on the lightray
($M=\mathbb{R}$) with $\alpha_{t}(\cdot)$ implementing the linear translation
on $\mathbb{R}$
\begin{align}
\omega_{\beta}(A)  &  =\left(  \Omega_{\beta},A\Omega_{\beta}\right)  ,\qquad
A,B\in\mathcal{A}(\mathbb{R})\\
F_{A,B}(t)  &  =\omega_{\beta}(\alpha_{t}(A)B),\quad F_{A,B}(t+i\beta
)=\omega_{\beta}(B\alpha_{t}(A))
\end{align}
Here the first line denotes the content of the GNS construction which
associates to a state on a $C^{\ast}$ algebra a concrete operator algebra
acting cyclically on a vector $\Omega_{\beta}$ in a Hilbert space (where in
the usual physicists manner we retain the same notation for the abstract
operator $A$ and its concrete Hilbert space representations). The second line
is the definition of the KMS property of the state $\omega_{\beta}$ which
consists in the existence of an analytic function in the strip $0<\mathrm{Im}%
\,z<\beta$ for every pair of operators $A,B\in\mathcal{A}(\mathbb{R})$ with
the stated boundary values. Assuming that the ground state theory exists, one
knows sufficient conditions under which the existence of the KMS state
associated with the time translation follows; these criteria are related to
quantum field theoretical phase space properties. In the case of chiral theory
these properties have the simple Gibbs form $\mathrm{Tr}\, e^{-\beta L_{0}%
}<\infty$ where $L_{0}$ is the standard notation for the rotational generator
in the 3-parametric M\"obius group. The theorem which geometrizes the abstract
thermal commutant of the KMS representation of $\mathcal{A}(\mathbb{R})$ reads

\begin{theorem}
(\cite{S-W}) The operator algebra associated with the heat bath representation
of $\mathcal{A}(\mathbb{R})$ at temperature $\beta=2\pi$ is isomorphic to the
vacuum representation restricted to the half-line chiral algebra such that
\begin{align}
(\mathcal{A}(\mathbb{R}),\Omega_{2\pi})  &  \cong(\mathcal{A}(\mathbb{R}%
_{+}),\Omega)\\
(\mathcal{A}(\mathbb{R})^{\prime},\Omega_{2\pi})  &  \cong(\mathcal{A}%
(\mathbb{R}_{-}),\Omega) \label{map}%
\end{align}
The isomorphism intertwines the translations of $\mathbb{R}$ with the
dilations of $\mathbb{R}_{+}$, such that also
\begin{align}
(\mathcal{A}((a,b)),\Omega_{2\pi})  &  \cong(\mathcal{A}((e^{a},e^{b}%
)),\Omega)
\end{align}

\end{theorem}

For the validity of this assertion it is important to be aware of the fact
that (different from the ground state representation of the global algebra
which as all global vacuum representations are always of quantum mechanical
type I$_{\infty}$) global KMS representations are of the same type as
restricted (localized) vacuum representations which are of hyperfinite type
III$_{1}$ i.e. equal to the \textit{monade} (appendix)$.$

The temperature $\beta=2\pi$ in Theorem 2 is in fact just a convenient choice.
For any other value, the exponential parametrization $e^{a}$ would change into
$e^{2\pi a/\beta}$. In the two dimensional version the plane is mapped into
the forward light cone and the abstract thermal commutant is mapped onto the
algebra of the backward light cone. The computational part of these
generalizations can be found in \cite{Bo-Yn}\cite{Bor}.

By lightfront holography, the area density of the localization entropy of a
wedge region is the localization entropy of the half-axis $\mathbb{R}_{+}$ in
an associated chiral CFT. By Theorem 2, the latter equals the heat bath
thermal entropy $S_{2\pi}$ of the global algebra $\mathcal{A}_{\mathrm{CFT}%
}(\mathbb{R})$ at temperature $\beta=2\pi$:
\begin{align}
S_{\mathrm{loc}}(A\times\mathbb{R}_{+})  &  =\vert A\vert\times S_{2\pi}%
\end{align}
Since the interval $I=(x,y)\subset\mathbb{R}_{+}$ is mapped under the
isomorphism of Theorem 2 to the interval $\log I= (\ln x,\ln y) \subset
\mathbb{R}$, and we expect that the heat bath entropy diverges proportional to
the length $L$ of an interval $\subset\mathbb{R}$, we get the finite
localization entropy
\begin{align}
S_{\mathrm{loc}}(A\times I)  &  \sim\vert A\vert\times\left|  \ln
\varepsilon\right|
\end{align}
where $\varepsilon= x/y\to0$. The exponentially parametrized thermodynamic
length factor $L$ is converted into an apparent short distance singular
behavior in $\varepsilon$.

There remains the calculation of the constant of proportionality, which will
be carried out in the next section by methods of chiral conformal QFT.

One note of caution: thermodynamic KMS states on the original massive bulk
matter have no simple relation to the massless case; they are belonging to
different theories (apart from the case where the original bulk matter is
conformal). It is only through the matter substrate-maintaining holographic
encoding that the chiral conformal theory enters the discussion. The remaining
question to what extent the $\varepsilon$ has an intrinsic meaning (i.e.\ with
an interpretation which is not added on but comes from the theory itself) will
be commented on in the section and in the appendix.

\section{Modular temperature-duality and the leading behavior of localization
entropy}

The isomorphism given by Theorem 2 plays the crucial role in mapping the
thermodynamic type I$_{\infty}$ limit sequence of the Gibbs system in an
increasing box into a type I$_{\infty}$ sequence which pictorially speaking
approximates the semi-axis from the inside. The appealing aspect of this
approximation is that, whereas in higher dimensional massive theories
different boxes correspond to different quantizations within basically the
same theory\footnote{Quantization boxes of different sizes define different
C$^{\ast}$ algebras even though ``morally'' they belong to the same system.},
chiral QFTs offer to do this within the same C$^{\ast}$ algebra. Let us
explain this point.

To get a finite localization entropy, we shall approximate the (equivalent)
thermal heat bath Gibbs state with respect to the generator $H$ of the
translations, by replacing $H$ with
\begin{align}
L_{0,R} = H + \frac1{4R^{2}}K,\qquad R\to\infty,
\end{align}
where $K=IHI$ is the generator of special conformal transformations, obtained
from $H$ by the conformal reflection $I$. $L_{0,R}$ is unitarily equivalent to
$(H+K)/2R = L_{0}/R$ by conjugation with a scale transformation $U(2R)$.
Because the unitary conjugation does not affect the trace and hence the
entropy, working with $L_{0,R}$ is the same as working with $L_{0}/R$. In
other words, we approximate $H$ by an operator with discrete spectrum, which
for $R\rightarrow\infty$ becomes dense and converges against the continuum of
the translation Hamiltonian.

The physical interpretation of this approximation is as follows: Because the
partition function $\,\mathrm{Tr}\, e^{-\beta L_{0}}$ can be regarded as the
partition function of the CFT in a ``box'' of size $2\pi$ (i.e.\ the circle
$S^{1}$), scaling $L_{0}$ by a factor $R$ amounts to passing to a box of
extension $L=2\pi R$.

In a very interesting recent paper \cite{Ni-To} it was shown that such a
``relativistic box'' interpolation is always possible for conformal theories
in arbitrary spacetime dimensions and that it is deeply related to Irving
Segal's attempt to use the Dirac-Weyl compactification of Minkowski spacetime
for cosmological purposes. In the $n$-dimensional case $H$ is the zero
component of the energy-momentum operator and $K$ the zero component of its
conformal reflected counterpart.

In the previous section it was shown that for chiral theories the global heat
bath thermal entropy at KMS temperature $\beta=2\pi$ and the localization
entropy for an (arbitrary) interval are two sides of the same coin; the only
difference is the parametrization which changes from a long distance $L
\rightarrow\infty$ (the size of the heat bath system) via $\varepsilon=e^{-L}$
to short distances (the size of the ``halo'' of the approximating interval
within $\mathbb{R}_{+}$). Hence the object which remains to be computed is the
partition function
\begin{equation}
Z_{0}(\beta) \equiv\,\mathrm{Tr}\,|_{H_{0}}e^{-\beta L_{0}} \quad
\hbox{at}\quad\beta= 2\pi/R = 4\pi^{2}/L.
\end{equation}
From this partition function the entropy follows in the standard way%
\begin{align}
S(\beta) =-\,\mathrm{Tr}\,\rho\ln\rho=\left(  1-\beta\partial_{\beta}\right)
\ln Z_{0}(\beta).
\end{align}
The remainder of the computation is done with the help of the temperature
duality relation which maps the partition function for large temperature into
one with small temperature (large $\beta$).

We assume that the chiral theory which appears in the holographic projection
is ``rational'', i.e.\ its observable algebra only admits a finite number of
unitarily inequivalent representations with associated representation spaces
$H_{\mu}$. According to Cardy and Verlinde this duality relation holds for the
partition functions with an appropriately shifted $L_{0}$
\begin{align}
&  \hat{Z}_{\mu}(\beta)\equiv\,\mathrm{Tr}\,|_{H_{\mu}}e^{-\beta({L}_{0}-\frac
c{24})} = e^{\frac c{24}\beta}Z_{\mu}(\beta),\nonumber\\
&  \hat{Z}_{\mu}(\beta)= \sum_{\nu}S_{\mu\nu}\hat{Z}_{\nu}(\frac{4\pi^{2}%
}{\beta}).
\end{align}
Relations of this kind first emerged from the Kac-Peterson study of characters
of loop groups, and geometrical structural arguments in favor of their general
validity for rational chiral models were proposed by Verlinde. The Verlinde
matrix $S_{\alpha\gamma}$ which appears in these relations is a priori not the
same as Rehren's ``statistics character''\footnote{Another invariant
equivalent definition is in terms of global invariant charges in the universal
C$^{\ast}$-algebra \cite{FRS}.} \cite{Re} (although both diagonalize the
system of fusion matrices), i.e.\ a numerical matrix related to the braid
group statistics data. There exists however a derivation based on modular
operator theory (see below) which shows that this is the case.

The remaining limit calculation $\beta\rightarrow\infty$ is almost trivial
since the leading term for $Z_{\mu}$ comes solely from the numerical $\frac
{c}{24}$ contribution, giving $Z_{\mu}(\beta) \approx S_{\mu0} e^{\frac{c}%
{24}\frac{4\pi^{2}}\beta}$ and thus
\begin{align}
S\left(  \beta=\frac{2\pi}R=\frac{4\pi^{2}}L\right)  \approx\frac{c\pi}6 R =
\frac{c}{12}L.
\end{align}
The charged sectors $\mu\neq0$ contribute to the entropy in the vacuum sector
only by nonleading terms, and the entropy in the charged sectors differs from
the entropy in the vacuum sector also only in the nonleading terms. In fact
the constant term in $o(\varepsilon)$ in the entropy in the sector $\mu$ turns
out to be $\ln S_{\mu0}$ \cite{Ka-Lo}. Hence the holographic matter content
enters the leading entropy term only through its \textit{algebraic structure}
and has no dependence on the superselected charges. Rewriting everything in
terms of the $\varepsilon$-parametrization one obtains the result of the
introduction (\ref{area}).

The best way to understand the temperature duality relation in an operator
setting is to view the partition function as the zero-point correlation
function in a unnormalized thermal state and to use modular theory in order to
perform an \textit{angular Euclideanization} \cite{S3}. The crucial formula is
the identity\footnote{This formula was derived in collaboration with Wiesbrock
\cite{S5}, but its physical role in Euclideanization was not explored.}
\begin{equation}
e^{-2\pi\tau L_{0}}=\Delta^{\frac{1}{4}}\tilde{\Delta}^{i\tau} \Delta
^{-\frac{1}{4}}.
\end{equation}
where $\Delta^{it}$ and $\tilde\Delta^{i\tau}$ are the modular groups
associated with the vacuum state and the algebras $\mathcal{A}(\mathbb{R}%
_{+})$ and $\mathcal{A}((-1,1))$, respectively. This can be interpreted as a
``modular Euclideanization'' as follows: The rotations by an imaginary angle
$2\pi i\tau$ are related to the M\"obius transformations which preserve the
interval $(-1,1)$, and hence act on $\mathbb{R}_{+}$ as two-sided compressions
for $\tau>0$. They become unitary operators if one changes the scalar product
of the Hilbert space with the ``metric'' $\Delta^{-\frac12}$.

Let us now define the rotational Gibbs state correlation functions of chiral
fields $\Phi_{k}$
\begin{equation}
\left\langle \Phi(\tau_{1},...\tau_{n});\beta\right\rangle _{\mu}%
\equiv\,\mathrm{Tr}\,|_{H_{\mu}}e^{-\beta({L}_{0}-\frac c{24})}\prod_{k}
\left(  e^{i2\pi\tau_{k}L_{0}}\Phi_{k}(0) e^{-i2\pi\tau_{k}L_{0}}\right)  .
\end{equation}
Then one finds the temperature duality relation for correlation functions
\cite{S3}
\begin{align}
\left\langle \Phi(i\tau_{1},\ldots,i\tau_{n});\beta\right\rangle _{\mu}  &
=\left(  \frac{2\pi i}{\beta}\right)  ^{a}\sum_{\nu}S_{\mu\nu}\left\langle
\Phi\left(  \frac{2\pi}{\beta}\tau_{1},\ldots,\frac{2\pi}{\beta}\tau
_{n}\right)  ;\frac{4\pi^{2}}{\beta}\right\rangle _{\nu}\\
a  &  =\sum_{i}\dim\Phi_{i}\nonumber
\end{align}
relating the ``Euclidean'' correlation functions (analytically continued to
imaginary angles) with correlation functions at the dual temperature.

For the multicomponent abelian current models, for which the $n$-point
functions functions can be expressed in terms of the Jakobi $\Theta
$-functions, these modular identities follow from known properties of $\Theta
$-functions \cite{B-M-T}\cite{S3}. The general (structural) derivation is
conceptually quite demanding; it can be found in \cite{S3}\cite{po-int}.

For more general chiral models beyond minimal models the temperature may not
be the only parameter which enters the description of thermal behavior. In
theories with a rich charge structure one may need the chiral analog of
chemical potentials.

Note that the localization entropy in the $\varepsilon\rightarrow0$ limit of
the auxiliary chiral theory does not depend on the length of an interval. In
the presence of several intervals corresponding to stochastically independent
systems, the partition functions factorized and the entropy is simply as
expected the sum of the contributions from the individual intervals.

Some critical remarks about our calculation of localization entropy are in
order. The existence of a conformal Hamiltonian with discrete spectrum permits
to replace the extrinsic quantization boxes used in the standard thermodynamic
limit by intrinsic relativistic boxes, i.e.\ sequences of states on the same
C$^{\ast}$ algebra. But the picture that the smaller relativistic boxes are
sitting inside the bigger ones is still somewhat metaphoric. This is also a
shortcoming of the usual thermodynamic limit $V\rightarrow\infty$ approach; in
fact the desire to formulate the thermodynamic limit in a more autonomous
manner was the main motivation for formulating statistical mechanics of open
sytems \cite{Haag}\cite{Bu-Ju}. It has been known for some time that the
so-called split property allows within the QFT setting to construct localized
thermodynamic limit sequences as well as their analogs for localization caused
thermal aspects. These are so-called ``funnel sequences'' (mentioned in the
introduction) of increasing type I$_{\infty}$ algebras $\mathcal{N}_{i}$ which
converge against a monade%
\begin{equation}
\mathcal{N}_{1}\subset\mathcal{N}_{2}\subset...\subset\mathcal{A}%
\end{equation}
In fact the field theoretic setting permits to construct a continuous sequence
$\mathcal{N}_{\varepsilon}$ where the type I$_{\infty}$ algebra $\mathcal{N}%
_{\varepsilon}$ is a canonically associated intermediate algebra between a
pair of monades $\mathcal{A(O}_{\varepsilon})\subset\mathcal{A(O})$ where
$\mathcal{O}_{\varepsilon}\subset\mathcal{O}$ are causally complete spacetime
regions (with the larger one having a nontrivial causal complement
$\mathcal{O}^{\prime})$ such that $\varepsilon$ measures the distance (minimal
spacelike distance) between the smaller inside the bigger region. The
restriction of the vacuum to $\mathcal{N}_{\varepsilon}$ turns out to be a
thermal Gibbs-like state. In the case of chiral theory $\varepsilon$ is simply
related to the minimal distance of a smaller interval from the two endpoints
of the bigger. Certain aspects of such a funnel approximation are implicit in
the work of Buchholz and Junglas \cite{Bu-Ju} on proving the existence of KMS
temperature states from the assumption that one knows the vacuum
representation. The \textit{split method is obligatory if one wants to compute
the energy or entropy for a finite split distance}, whereas the above
relativistic box method is expected to account only for the leading term. This
also implies that the re-interpretation of the $L_{0}$ temperature in terms of
a geometric box of size $2\pi R$ (common among string theory users of
conformal QFT) only looses its metaphorical character in the limit
$R\rightarrow\infty$; only the more ambitious splitting procedure is the
autonomous method to construct type I$_{\infty}$ subalgebras which are
localized in a subregion. Contrary to a momentum cut-off which changes the
theory in a conceptually uncontrolled fashion\footnote{A momentum space
cut-off is an extremely ill-defined concept. Even in those cases in which the
local theory (e.g. through its form factors or correlation functions) is
explicitly known (example chiral theories, factorizing models), nobody has an
idea by what controllable manipulation on the local theory one can obtain a
mathematically well-defined and physically interpretable cutoff model.}, the
split property creates a physical distance $\varepsilon$ \textit{within a
given local theory}. But since a local theory has no elementary length,
$\varepsilon$ is not fixed by the theory. Needless to say that setting it
equal to the Planck length, leads (apart from the dependence on the parameter
$c$ associated to the holographic quantum matter) to the entropy Bekenstein
entopy formula.

The problem with this totally intrinsic split method is that it is easy to
show the existence of a funnel approximation \cite{Do-Lo} but it has turned
out to be extremely hard to do computations. One expects that the description
of this method (for more details see the appendix) may be important for the
future development.

\section{Concluding remarks}

Thermal aspects caused by the quantum field theoretic vacuum polarization at
boundaries of causally complete localization regions are in several aspects
different from the classical heat bath thermal behavior. In the case of a
wedge region one finds that the vacuum polarization leads to an area law for
the entropy where the area refers to the edge of the wedge. The conformal
invariance of the lightfront projection reveals however an unexpected relation
between a global heat bath thermal system at a fixed temperature (in our case
$\beta=2\pi$) and the thermal aspects of a system caused by vacuum
fluctuations as a result of localization in the direction of the lightray. The
result is a one-parametric family of area densities for the entropy which in
the limit of $\varepsilon\rightarrow0$ approach a universal (independent of
the approximating sequence) $\vert\ln\varepsilon\vert$ behavior; the parameter
$\varepsilon$ measures the size of the ``vacuum polarization halo'' which, in
agreement with Heisenberg's observation about the boundary nature of vacuum
polarizations, diverges in a universal manner with a strength which only
depends on the holographic matter content.

Our result shows in particular that the conceptual basis of previous
calculations of localization entropy as entanglement entropy associated with
the energy levels of the standard time translation Hamiltonian \cite{Bo-So} is
not sustainable. In those calculations a free scalar field is restricted to
the exterior of a sphere (the model of a black hole) by simply factorizing the
quantum mechanical energy levels into their inside/outside contribution. The
infinity caused by vacuum fluctuations as a result of the sharp factorization
is parametrized in terms of a model-changing momentum space cutoff.
Choosing the latter of the order of the Planck scale the numerical result is
consistent with the quantum interpretation of Bekenstein's classical area
formula. However the calculation has hardly anything to do with localization
entropy and the idea of a vacuum polarization halo at the causal horizon is at
best a metaphorical interpretation but not an autonomous property. Most of the
quantum entropy calculations are in fact based on a counting picture where
energy levels are populated and the occupied levels are counted. This kind of
quantum mechanical picture of entropy is at odds with the recently discovered
principle of quantum local covariance of QFT in curved spacetime (in
particular in Minkowski spacetime) which is also related to the background
independence \cite{Br-Fr}. Such level-counting calculations of energy and
entropy have been used in the calculation of the cosmological constant as well
as in string theory based microscopic calculations of entropy of certain
limiting cases of black holes. According to a recent paper by Hollands and
Wald \cite{Ho-Wa} entitled: ``\textit{Quantum Field Theory is not merely
Quantum Mechanics applied to low energy effective degrees of freedom}'' such
computations should be looked upon with suspicion because they violate one of
the most cherished principles underlying general relativity. Since both
classical relativity and local quantum physics should come together in a
future theory of QG, the unknown principles of QG should constitute a
synthesis and not a negation of the known principles. Modular localization, on
which the present considerations were based, is very different from filling
levels in momentum space and, if one succeeds to extend modular theory to QFT
in curved spacetime (modular origin of diffeomorphisms), will be fully
consistent with local covariance and background independence.

Of course the physically realistic treatment of the Hawking effect for a
collapsing star cannot be given in terms of equilibrium KMS states on
localized algebras (e.g. the algebra outside the Schwarzschild radius). But as
the historical calculations in such a nonequilibrium situation show
\cite{Haw}\cite{Fr-Ha}, one needs only local information in the infinitesimal
neighbourhood of the formation of the black hole horizon in order to
understand the thermal nature of the outgoing asymptotic radiation. Recently
the conceptual framework for statistical mechanics entropy has been enlarged
to include stationary nonequilibrium states where entropy flows from one into
another changing state. There can be little doubt that a realistic black hole
entropy discussion must take these new concepts into consideration and that a
future understanding of black hole entropy will have to find its place in a
future theory of nonequilibrium statistical mechanics.

Acknowledgement: I am indebted to K.-H. Rehren for a critical reading which
led to essential improvements in the presentation.

\appendix

\section{Some facts about modular operator theory}

For the convenience of the reader we mention some mathematical concepts
concerning modular aspects of operator algebras \cite{Su} which have been
freely used in the main text. Their application requires the algebraic
formulation of QFT in the sense of Haag's book on \textit{Local Quantum
Physics} \cite{Haag}.

Modular theory associates to a \textquotedblleft standard\textquotedblright%
\ pair ($\mathcal{A},\Omega$) of an operator in a Hilbert space and a vector
$\Omega$ on which it acts in a cyclic and separating (the only annihilator of
$\Omega$ in $\mathcal{A}$ is the zero operator) way a (one-parametric) unitary
modular group $\Delta^{it}$ and an antiunitary idempotent operator $J$. These
\textquotedblleft modular data\textquotedblright\ result from the densely
defined closable antilinear Tomita S-operator by polar decomposition according
to
\[
S:A\Omega\mapsto A^{\ast}\Omega,\qquad\Delta:=S^{\ast}S,\qquad J:=S\Delta
^{-\frac{1}{2}}=J^{-1}.
\]
The prime statements of the Tomita-Takesaki theory are the relations
\[
\sigma_{t}(\mathcal{A}):=\mathrm{Ad}_{\Delta^{it}}\mathcal{A}=\mathcal{A}%
,\qquad J\mathcal{A}J=\mathcal{A}^{\prime};
\]
the adjoint action of the modular unitary defines the \textit{modular
automorphism} of $\mathcal{A}$ and the modular inversion $J$ defines an
antiunitary isomorphism onto the commutant (thus showing that in a standard
situation the algebra is anti-isomorphic with its commutant, which excludes
the irreducibility of a standard situation). Certain aspects of the spectrum
of the modular group determine the \textquotedblleft type\textquotedblright%
\ (isomorphism class) of the operator algebra; in particular for type
III$_{1}$ the spectrum of the infinitesimal generator is purely continuous (in
fact, $=\mathbb{R}$.)

From case studies and general structural arguments one knows that the local
algebras of QFT are isomorphic to the unique hyperfinite type III$_{1}$ factor
von Neumann algebra which (specifically in this quantum field theoretic
setting) for brevity as well for more profound reasons (see introduction) is
referred to as a \textit{monade}; they are standard with respect to the vacuum
(the Reeh-Schlieder property). The global algebras of QFT on the other hand
are of type I$_{\infty}$, but they loose this quantum mechanical property in
thermal states; in this case they acquire the same algebraic structure as the
local algebras, namely hyperfinite type III$_{1}.$ Whereas the commutant in
the heat-bath thermal situation remains an abstract thermal shadow world, the
commutant in the localized vacuum situation is geometric and in typical cases
equal to the algebra of the causal disjoint (Haag duality). The only case in
which the modular group acts geometrically (independent of the particular
model of QFT) is the wedge situation ($\mathcal{A}(W),\Omega_{res}$); there
are however many ``partially geometric'' situations in which the reference
state is different from the vacuum and the corresponding modular group acts as
a diffeomorphism if restricted to the subalgebra \cite{S3}.

The modular theory was significantly enriched by the concept of modular
inclusion and of modular intersection \cite{Wi}\cite{Bo}. These structures are
related to a generalization of the Takesaki theorem in section 2: instead of
requiring that the modular group of the larger algebra acts as a
one-parametric automorphism group on the smaller, one only assumes that it
contracts the algebra in one direction ($\pm$ halfsided modular inclusions).
These structures can then be used to show, that a QFT with all its structural
richness including its covariances and geometric and spectral aspects emerges
from the pure algebraic modular positioning of a finite number of copies of
representations of the monade in a joint Hilbert space \cite{Ka-Wi}. This view
seems to be very powerful for a better understanding about the relation
between algebraic properties, geometry and thermal aspects and may well lead
to a third path towards quantum gravity (a modular path). It underlies the
work of Vaughn Jones who generalized the symmetry concept behind compact
groups; for his purpose (which does not include the incorporation of thermal
KMS and modular localization aspects) he was able to work with different
monades namely the unique hyperfinite type II$_{1}$ algebra \cite{Jones},
which as a result of its tracial state is a more amenable mathematical object.

A closely related modular concept which permits to implement many ideas (which
in the Lagrangian approach required to imagine momentum space cutoff)
within the given local QFT, is the so called \textit{split property}. Although
it will not be used in this article for computations (because it belongs to
those modular properties which still resist computational attempts), it
provides by far the best conceptual setting for localization entropy. Let us
finally close this section with some remarks on the split property and closely
related previous attempts to generalize the notion of entropy beyond the
time-honored von Neumann entropy definition.

There are several candidates for such a definition with a similar
physical-intuitive content. One attempt employs the framework of the
Connes-Narnhofer-Thirring entropy \cite{C-N-T}\cite{Na} which is a kind of
relative entropy \cite{Araki}\cite{Kosaki}; it associates (adapted for the
present purpose) to an inclusion of two algebras and a state $\omega$ denoted
as ($\mathcal{A}\subset\mathcal{B},\omega$) an entropy $H_{\mathcal{B}}%
(\omega,\mathcal{A})$. The definition is such that in case of $\mathcal{A}%
=\mathcal{B}$ being quantum mechanical type I$_{\infty}$ it agrees with the
von Neumann entropy where $\omega$ is represented by a density matrix. A
closely related idea which is based on the more restrictive assumption of a
``split'' inclusion ($\mathcal{A} \subset\mathcal{B},\omega$) is due to
Doplicher and Longo \cite{Do-Lo}. This split property gives rise to a
functorial association of an intermediate type I$_{\infty}$ algebra
$\mathcal{N}$ (which can be explicitly written in terms of modular data
\cite{Do-Lo})
\begin{align}
\label{A 2}\mathcal{A}  &  \subset\mathcal{N\subset B}.
\end{align}
The vacuum state $\omega$ restricted to the type I$_{\infty}$ algebra
$\mathcal{N}$ is a density matrix $\rho(\mathcal{N},\omega)$ (in terms of the
tracial weight formalism) to which the von Neumann definition of the entropy
may be applied (\cite{Do-Lo} page 511).

In the case of a pair of localized algebras $\mathcal{A}=\mathcal{A}%
(\mathcal{O})$, $\mathcal{B}=\mathcal{A}(\hat{\mathcal{O}})$, $\mathcal{O}%
\subset\hat{\mathcal{O}}$ in local quantum physics, the split property later
turned out to be the consequence of a physical phase space degree of freedom
behavior \cite{A-B-F} called ``nuclearity''. In this case the intermediate
algebra $\mathcal{N}$ may be thought of as a quantum mechanical system of
``fuzzy'' localization between $\mathcal{O}$ and $\hat{\mathcal{O}}$, allowing
the vacuum-polarization ``halo'' to thin out softly. The intuitive content of
the halo parallels the transition region of the test functions with which one
smears charge densities of conserved currents in order to define partial charges.

The split property is intimately linked with the notion of correlation-free
product states. The functorial construction of the intermediate typ
I$_{\infty}$ algebra starts from the assumption that the uncorrelated factor
state (the split state)
\begin{equation}
\omega_{\mathrm{split}}(AB^{\prime}):=\omega(A)\omega(B^{\prime}%
)\quad\hbox{for}\quad A\in\mathcal{A}(\mathcal{O}),B^{\prime}\in
\mathcal{B}^{\prime}=\mathcal{A}(\hat{\mathcal{O}})^{\prime}\label{A 3}%
\end{equation}
is a normal state\footnote{This assumption is guaranteed in QFT by the
nuclearity property \cite{A-B-F}.} (i.e.\ it has natural continuity properties
with respect to the involved operator algebras); such a state according to
modular theory possesses a distinguished vector representative $\eta
\in\mathcal{P}(\mathcal{A\vee B}^{\prime},\Omega)$ in the natural cone
associated with the algebra generated by $\mathcal{A}$ and $\mathcal{B}%
^{\prime}$ and the vacuum, i.e.\ $\omega_{\mathrm{split}}(AB^{\prime
})=\left\langle \eta\left\vert AB^{\prime}\right\vert \eta\right\rangle $. The
properties of this vector lead to the unitary equivalence $W$ of the vacuum
representation of the algebra $\mathcal{A\vee B}^{\prime}$ with the tensor
product representation $\mathcal{A\otimes B}^{\prime}$ on $H\otimes H$; the
intermediate type I$_{\infty}$ factor algebra $\mathcal{N}$ turns out to be
simply $W(B(H)\otimes1)W^{\ast}$, while $P(\mathcal{O},\hat{\mathcal{O}%
})=W(1\otimes P_{\Omega})$ is the projector onto the factor space
$\overline{N\left\vert \eta\right\rangle }$.

Here we used local commutativity of $\mathcal{A}$ and $\mathcal{B}^{\prime}$
algebras in order to arrive at stochastic independence in the sense of
existence of correlation-free product states. It is interesting to note that
vice-versa, with a stronger notion of absence of correlation \cite{Bu-Su} one
is able to characterize the commutativity of algebras in terms of existence of
correlation-free states. This shows that local commutativity is inexorably
linked with stochastic independence for causally disjoint observation,
i.e.\ that some form of relativistic causality is not an option.

The split construction is very important to reconcile a KMS localization
temperature with a finite localization entropy. Strictly speaking the spatial
interpretation of the thermodynamic limit sequence (and the related analogous
inner exhaustion of the vacuum state restricted to a sharply localized algebra
by a sequence of type I$_{\infty}$ algebras) is ``metaphorical''. But the
split property allows to replace this argument by a completely autonomous one
in which the box sequence is replaced by a sequence of fuzzily localized type
I$_{\infty}$ algebras, forming a genuine inclusive exhausting sequence
(``funnel'') inside the ``monade'' which describes the open thermal system.

\end{document}